# SERVICE DISCOVERY – A SURVEY AND COMPARISON

Bendaoud Karim Talal[1] and Merzougui Rachid[2]

[1] STIC Laboratory, Department of Telecommunication, University of Tlemcen, Algeria
bendaoud.talal@gmail.com
[2] STIC Laboratory, Department of Telecommunication, University of Tlemcen, Algeria
merzrachid@yahoo.fr

## ABSTRACT

*With the increasing number of services in the internet, companies' intranets, and home networks: service discovery becomes an integral part of modern networked system. This paper provides a comprehensive survey of major solutions for service discovery. We cover techniques and features used in existing systems. Although a few survey articles have been published on this object, our contribution focuses on comparing and analyzing surveyed solutions according eight prime criteria, which we have defined before. This comparison will be helpful to determine limits of existing discovery protocols and identify future research opportunities in service discovery.*

## 1. INTRODUCTION

With the increasing number of services in the internet, companies' intranets, and home networks service discovery becomes an integral part of modern networked system. This process is simple if the user and service provider know each other at run time [1], so what is needed is an efficient mechanism which ensures high availability of services to users and applications, and high utilization of services. In this article we survey a number of service discovery approaches. Despite the existence of a number of survey papers regarding service discovery protocols [2-5], we believe that a comprehensive overview of techniques and open issues for service discovery is useful. The purpose of this article is to provide a comprehensive review on service discovery approaches and to compare and analyze surveyed solutions according eight prime criteria. In this manner, we use the work both as a survey and as a guide for the design of a service discovery system.

The structure of the paper is as follow: section 1 defines the objective, features and techniques of existing service discovery. Based on that, section 2 provides a comprehensive survey for leading technologies in this area. Service discovery protocols discussed in section 2 will be then compared taking various criteria in section 3. Section 4 concludes this paper with a list of future research opportunities in service discovery. In Appendix 1, we present the summary of our comparison

### 1.1. Service Discovery Definition [6]:

Service discovery provides a mechanism which allows automatic detection of services offered by any node in the network. In other words, service discovery is the action of finding a service provider for a requested service. When the location of the demanded service is retrieved, the user may further access and use it.

## 1.2. Service Discovery: Objectives, Features, and Techniques:

According to[6][7], the objective of a service discovery mechanism is to develop a highly dynamic infrastructure where users would be able to seek particular services of interest, and service providers offering those services would be able to announce and advertise their capabilities to the network . Furthermore, service discovery minimize human intervention and allows the network to be self-healing by automatic detection of services which have become unavailable. Once services have been discovered, devices in the network could remotely control each other by adhering to some standard of communication.

In what follows we define the service discovery features and techniques that have been specified to achieve these objectives. A summary of service discovery features is given in figure 1.

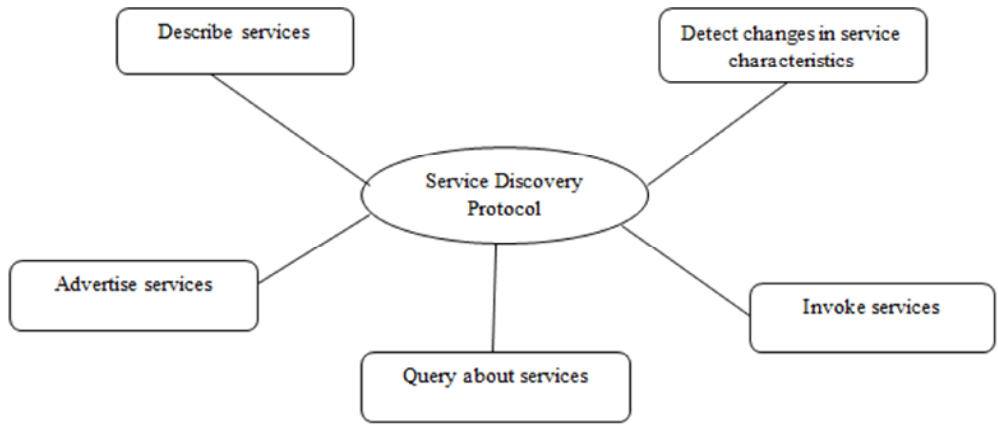

Figure-1- Service Discovery Features

## 1.2.1. Service Description:

In order to facilitate the service discovery process, each protocol has a description language to define the vocabulary and syntax used to describe the service and its properties [8]. The available methods for this task vary according to the degree of expressiveness: key/value, template-based and semantic description [1]. In the key/value approach, services are characterized using a set of Attribute-value pairs. The template-based approach: uses the same technique as in the first approach, in addition it offers predefined set of common attributes which are frequently used. The semantic description relies on the use of ontology. It has richer expressive power than the first two approaches.

## 12.2. Service Discovery Architecture:

Architecture used by service discovery protocols can be classified as directory and non-directory based models [4], according to how the service descriptions are stored.

***The directory based model*** has a dedicated directory which maintains the whole service descriptions. In this case, the directory takes care of registering service descriptions and processing user requests. The directory can be logically centralized but physically distributed over the network. Therefore, service descriptions are stored at different locations (directories).

***The non-directory based model***: has no dedicated directory, every service provider maintains its service descriptions. When a query arrives, every service provider processes it and replies if it matches the query.

### 1.2.3. Service Announcement and Query:

Service announcement and query are the two basic mechanisms for clients, service providers, and directories to exchange information about available services.

***Service Announcement***: allows service providers to indicate to all potential users that a set of new services is active and ready for use. This will be accomplished by registering the appropriate service descriptions with the directory if it exists, or multicast service advertisements.

***Query approach***: allows users to discover services that satisfy their requirement. To do this, users initiates (a) unicast query to the directory, or (b) multicast query [7]. The query is expressed using the description language [6], and specifies the details about service it is looking for. The directory or service provider that holds the matching service description replies to the query.

When a directory exists, service providers and users will first discover the directory location before services can be registered and queried. In this case, the directory can be seen as any service in the network and makes advertisement to advertise its existence.

### 1.2.4. Service Usage (Service invocation):

After retrieving the desired services information, the next step is to access. However, apart from performing service discovery, most protocols offer methods for using the services [6]. An example is Simple Object Access Protocol (SOAP) used in Universal Plug and Play (UPnP).

### 1.2.5. Configuration Update (management dynamicity)

Service discovery protocol must preserve a consistent view of the network and deliver valid information about available services while network is dynamic. Therefore, the management of such dynamicity is required. Configuration update allows users to monitor the services, their availability and changes in their attributes. There are two sub functions in Configuration Update:

- ***Configuration Purge*** Allows detection of disconnected entities through (a) *leasing* and (b) *advertisement time-to-live (TTL)* [7]. In leasing, the service provider requests and maintains a lease with the directory, and refreshes it periodically. The directory assumes that the service provider who fails to refresh its lease has left the system, and purges its information. With TTL, the user monitors the TTL on the advertisement of discovered services and assumes that the service has left the system if the service provider fails to re-advertise before its TTL expires.
- ***Consistency Maintenance:*** Allows users to be aware when services change their characteristics. Updates can be propagated using (a) push-based *update notification*, where users and directories receive notifications from the service provider, or (b) pull-based *polling for updates* by the user to the directory or service provider for a fresher service description [7].

It is important to note that the features and techniques mentioned before representing the pillars around which an autonomic service discovery protocol is based. But, depending on characteristics of each protocol other functions have been already proposed in diverse approaches (e.g. service selection, security, scalability).

# 2. Various Service Discovery Approaches Adopted by Industry

Over the past years, many organizations and major software vendors have designed and developed a large number of service discovery protocols. This section provides a comprehensive survey for leading technologies in this area and examines functional issue defined in the previous section for each protocol.

## 2.1. SLP:

Service location Protocol (SLP) [9] is an open, simple, extensible, and scalable standard for service discovery [10] developed by the IETF (Internet Engineering Task Force). It was intended to function within IP network. SLP addresses only service discovery and leaves service invocation unspecified [11]. The SLP architecture consists of three main components:

•**User Agent** (**UA**)**:** software entity that sends service discovery request on a user application's behalf.
•**Service Agent (SA):** advertises the location and characteristics of services on behalf of services.
•**Directory Agent (DA):** a central directory collects service descriptions received from SAs in its database and process discovery queries from UAs.

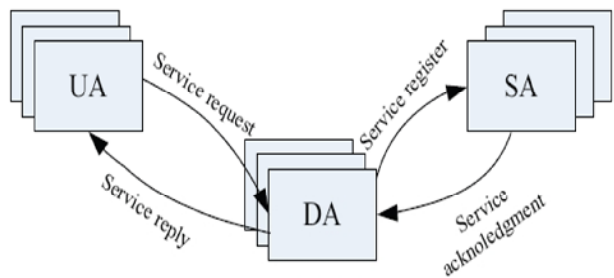

Figure -2- SLP agents and their transactions for service discovery and registration.

As shown in figure 2, when a new service connects to the network, the SA contacts the DA to advertise its existence (Service Registration). Registration message contains: service lifetime, URL for the service, and set of descriptive attributes for the service. Both URL schemas and attributes are defined in [12]. Registration should be refreshed periodically by the SA to indicate its continuous existence. The same when the user needs a certain service, the UA sends request message to the DA which in turn responds with message containing URLs for all services matched against the UA needs. The client can access one of the services pointed to by the returned URL. The protocol used between the client and the service is outside the scope of the SLP specification [13].

To perform their respective roles UA and SA have first to discover DA location. SLP provides three methods for DA discovery: static, active, and passive.

In static approach: SLP agents obtain the address of the DA using DHCP; the necessary DHCP options for SLP are defined in [14]. With active approach: SLP agent (UA/SA) sends service request to the SLP multicast group address, a DA listening on this address will respond via unicast to the requesting agent. In passive approach: DA multicasts advertisements periodically, UAs and SAs learn the DA address from the received advertisements.

It is important to note that the DA is not mandatory; it is used especially in large networks to enhance scalability. In smaller network (e.g. home network, office network) there may be no real need for DA, SLP is deployed without DA. In this case, UAs send their service requests to the SLP multicast address. The SAs announcing the service will send a unicast response to the UA. Moreover, SAs announce their presence via multicast

SLP provides a powerful filter that allows UAs to select the most appropriate service from among services on the network. The UA can formulate expressive queries using operators such as AND, OR, comparators (<, =,>, <=,>=) and substring [2].

SLP is an open source; it does not depend on any programming language and scales well in large networks. The scalability is supported by various features such as scope concept, and multiple DAs.

## 2.2. Jini

Jini [15] is a distributed service discovery system developed by Sun-Microsystems in java. The goal of the system is the federation of groups of clients/services within a dynamic computing system [1]. A Jini federation is a collection of autonomous devices which can become aware of one another and cooperate if need be.

To achieve this goal, Jini uses a set of lookup services to maintain dynamic information about available services and specifies how service discovery and service invocation is to be performed among Java-enabled devices [13]. The Jini discovery architecture is similar to that of SLP:

**Client:** requests Lookup Service for available service.
 **Service provider**: registers its services and their descriptions with Lookup Service.

**Lookup Service (LS)**: Directory which collects service descriptions and process match queries in manner analogous to DA in SLP. Unlike SLP, where DA is optional, Jini operates only as a directory based service discovery and requires the presence of one or more Lookup Services in the network.

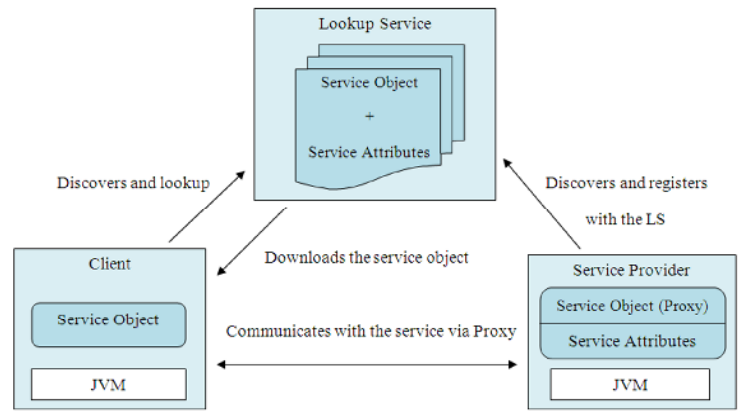

Figure-3- Jini Architecture

The heart of Jini is a trio protocols called: discovery, join, and lookup. Discovery occurs when a service provider or client is looking for Lookup Service. Join occurs when a service provider has located a LS and wishes to join it. Lookup occurs when the client needs to locate and invoke a service.

**Discovery:**
At startup, service providers and clients have first to discover one or more LSs in the network. This may be done either through the *multicast announcement protocol*, i.e. by listening to the multicasted LSs advertisements, or through *multicast request protocol,* i.e.by sending multicast requests for lookup services. Once LS is discovered, clients and service providers can use *Unicast Discovery Protocol* to directly communicate with the discovered LS.

**Join:**
Once an LS has been discovered, the service provider can advertise its service by registering a service object (proxy) and its attributes with the LS. The registration is lease based and service provider has to renew the lease to maintain its listing in the LS. The proxy contains the java interface for the service, and invocation methods that users will invoke to execute the service. This object can be the service itself or a smart object that offer direct access to the service over the network [2].

**Lookup:**
The Jini client requests the discovered LS for a particular service by its attributes. Whereas a service request in SLP returns the service URL, the LS returns one or more matching proxy objects. The client downloads a copy of the proxy, and then interacts directly with the service. The location of the service is unimportant from the client side, because the proxy object encapsulates the location and the protocol necessary to operate it.

Jini uses Java's remote method invocation (RMI) facility for all interactions between either a client or a service and the lookup server (after the initial discovery of the lookup server) [3]. It allows data as well as objects to be passed through the network.

In Jini, evaluation of requests is based on equality and exact correspondence between request parameters and attributes of services. Jini does not allow the evaluation of complex queries with Boolean operators or comparators such as SLP.

Configuration Update in Jini is guaranteed by using the two following concepts: Leasing and remote events.

***Leasing:*** The service provider maintains a lease with the LS, where it periodically refreshed. Leasing ensures that failing services get recognized and automatically removed from the Jini Lookup Service.

***Remote Events:*** allows Jini clients to register their interests in events of another object and can be notified whenever these events happen.

## 2.3. UPnP
UPnP [16] is a Microsoft-developed service discovery technology aimed at enabling the advertisement, discovery, and control of networked devices and services. It is built upon IP that is used for communication between devices, and uses standard protocols like HTTP, XML, and SOAP for discovery, description, and control of devices. The architecture of UPnP network is as follow:

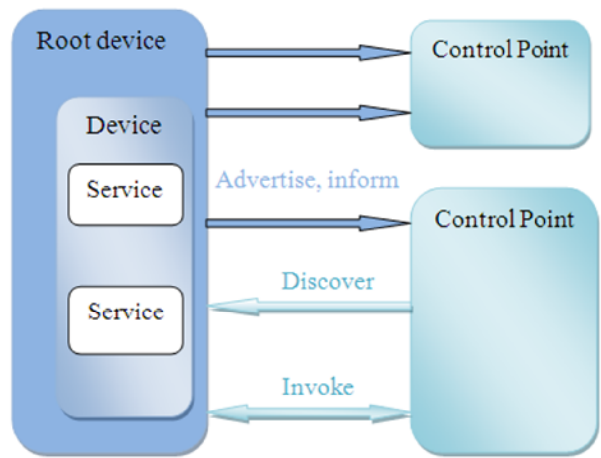

Figure-4- The UPnP architecture components.

**_Device:_** can be any entity on the network that contains services or any embedded devices. A service is the smallest unit of control in UPnP and it consists of:

- _State table:_ models the state of the services at run time through state variable.
- _Control server:_ receives requests, executes them; updates the state table and returns responses.
- _Event server:_ publishes events to interested clients when service state changes.

**_Control point:_** any entity in the network that is able to discover, retrieve service descriptions, and control the features offered by a device.

The UPnP features can be epitomized as the following five steps:

**_Addressing:_** When a device is connected to the network, it tries to get an IP address. UPnP defines two methods for that. The first method is requesting an IP address via DHCP. If no DHCP server is available, the IP address is determined by AUTO-IP. The device claims an address by randomly choosing an address in the reserved range and then making an ARP request to see if anyone else has already claimed that address.

**_Description_**: UPnP uses a non-directory based approach for service discovery where each device hosts a device description document. This document is expressed in XML and includes device information (e.g. manufacturer, model, serial number ...), list of any embedded devices or services, as well as URLs for the service description, control, and eventing. For each service, the description contains the service type, service ID, state table, and list of the actions that a service can perform.

**_Discovery_**: The UPnP discovery process is based on the Simple Service Discovery Protocol (SSDP). This protocol allows UPnP devices to announce their presence to others and discover other devices and services. When a device comes on-line, it sends advertisement (ssdp: alive) via multicast to announce its presence. The advertisement message is associated with a lifetime and contains typically the type of the advertised service, and URL to the description. An UPnP device may send out many presence announcements. When the device wish to disconnect from the network, it should send an advertisement (ssdp:bye-bye) to notify control points that its services are no longer available. Any control point that comes on-line after the UPnP device has announced its presence sends out discovery request (ssdp: discover) via multicast. Devices listening for this multicast respond via unicast if they match the service. Control points can search only for: all services, specific service type, or specific device type since SSDP does not support attribute-based querying for services [13].

***Control:*** Having retrieved the description of the device from the URL provided in the response message, the control point can invoke the service. To do this, the control point sends a suitable control message to the control URL for the service. Control messages are expressed in XML using the Simple Object Access Protocol.

***Eventing:*** Similarly to Jini, UPnP allows control points to register for and receive notifications of device state changes using eventing feature. The service publishes updates by sending event messages which contains list of state variable and their current values. These messages are also expressed in XML and formatted using the General Event Notification Architecture (GENA).

## 2.4. SALUTATION

Salutation [17] is another major cooperation architecture developed by the salutation consortium. It is an open standard independent of operating systems, physical platforms and communication protocols which addresses the problems of service discovery. One important difference of Salutation from other service discovery technologies is that it is transport-protocol independent.

As shown in figure-5-, Salutation architecture consists of two main components: Salutation Manager (SLM) and Transport Manager (TM). Each salutation-enabled device implements a local SLM which stores descriptions of local services. An SLM can find other SLMs in the network and discover services they offer. Communication between SLMs is based on remote procedure call (RPC). The Transport Manger (TM) acts as an interface between the SLM and the communication technology. The TM is introduced to isolate the implementation of the SLM from particular transport-layer protocols and thereby gives Salutation network transport independence [3]. One SLM may have many TMs in order to operate over different network technologies (e.g. IR, Bluetooth..)[13].

Salutation defines a specific format for describing services called: *Functional Unit Description Record (FUDR)*. This record consists on the service type and set of Attribute Records. Each Attribute Record contains identifier of the attribute, an identifier of the comparison function to be used to compare the value of the attribute in the event of request and finally the value of the attribute.

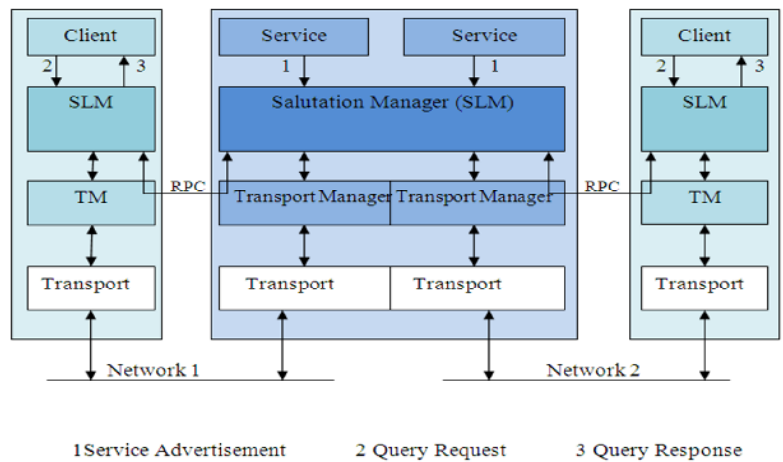

Figure-5- Salutation Architecture

A summary of the main tasks of the SLM follows:

**Service Registry**: Similar to SLP, Salutation can also function as both directory and non-directory discovery mechanism. The SLM contains a registry to hold information about services. The minimum requirement for the registry is to store information about services connected to its SLM. Optionally, the SLM registry may store information about services that are registered in other SLM. In this case this SLM would be designed as the central directory (in the case of local networks).

**Service Discovery:** A client can discover services by sending a request (*Query Capability Call*) to the local SLM which cooperates with others SLMs to interchange information about registered services and responds with a list of all matched FUDRs in reply. This cooperation among Salutation Managers forms a conceptually similar lookup service to Jini [2]. Remote services are discovered by matching type and set of attributes specified by local SLM.

**Service Session Management:** When a user wants to use a discovered service, it requests its SLM to establish a session with the service. Communication between clients and services can be in a number of ways:

- *Native mode***:** messages are exchanged through a native protocol. In this case, Salutation Manager is used solely to discover services of other network.
- *Emulated mode:* Salutation Manger is used to carry message exchange but does not inspect the contents. This is useful when common messaging protocol does not exist between clients and discovered services.
- *Salutation mode***:** message format and exchange protocol are defined by salutation Manager.

**Service Availability:** The Salutation Manager does not implement the mechanism of leasing or notification to control the availability of a service, but a client can periodically check the availability of a service by asking its local Salutation Manager. The local Salutation Manager requests the appropriate Salutation Manager to perform an Availability Check. The period of the Availability Check is specifiable.

## 2.5. Bluetooth SDP

Bluetooth [5] is a new short range wireless transmission technology, developed by the Bluetooth Special Interest Group. This technology allows Bluetooth-enabled devices to communicate via short-ranged radio links with low power and low cost. Bluetooth devices organize themselves into a personal area network called piconet. A piconet can at most consist of eight active devices whereas the member that initiates the communication becomes the master of the piconets. Groups of piconets communicating with each other are called scatternet.

Bluetooth Service Discovery Protocol (SDP) is service discovery protocol that is local to a single piconet [8]. It is used to locate services provided by Bluetooth devices. Contrary to other service discovery protocols, Bluetooth SDP offer limited functionality and addresses only discovery (It does not provide service advertisement, service usage, and there is no configuration purge mechanism) [13].

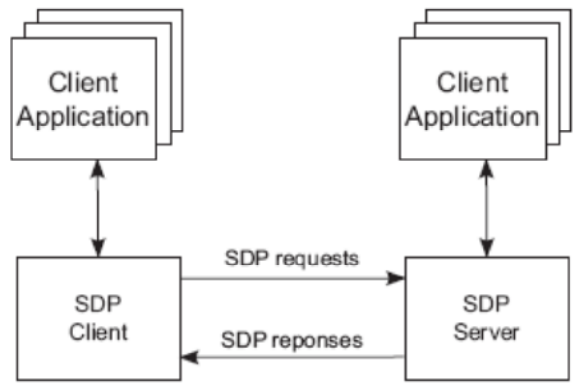

Figure-6- SDP client-server interaction

In Bluetooth, a service is any entity that can provide information or perform an action. Each service is described by a service record consisting of a set of attribute-value pairs. Information about services is stored in SDP servers. A Bluetooth device wanting to use a service is called SDP client. The Bluetooth SDP defines two modes for discovery: searching and browsing. In the first mode the SDP client can discover services by type and attributes, while browsing allows SDP client allows SDP client to get list of all services.

## 2.6. SECURE SERVICE DISCOVERY SERVICE (SSDS)

The Ninja project by the University of California, Berkeley developed the Service Discovery Service (SSDS) [18]. SSDS is a directory based mechanism for service deployment and discovery. It is similar in many respects to other service discovery protocols discussed. Like Jini, SSDS is implemented in Java and depends on Java RMI. Regarding service description, SSDS is based on XML like UPnP. This is a powerful combination given the expressiveness of XML and portability of Java. One important difference of SSDS from other service discovery technologies is that it features much stronger security and scalability than those discussed so far.

Besides services and clients, the SSDS architecture is composed of:

**SSDS server:** acts as Directory Agent in SLP, it maintains service description about available services and processes clients' queries. For the purpose of scalability, SSDS servers are organized into multiple shared hierarchies, so that tasks can be shared among several servers.

**Certificate Authority (CA):** the SSDS server uses certificates signed by the CA for the authenticity between the components of the systems.

**Capability Manager (CM):** the SSDS server uses capabilities (access rights) as a mechanism for controlling access to service descriptions. In other word, a capability proves that a particular client is allowed to access a particular service. The capability manager generates and distributes capabilities for all users who are entitled to use particular service.

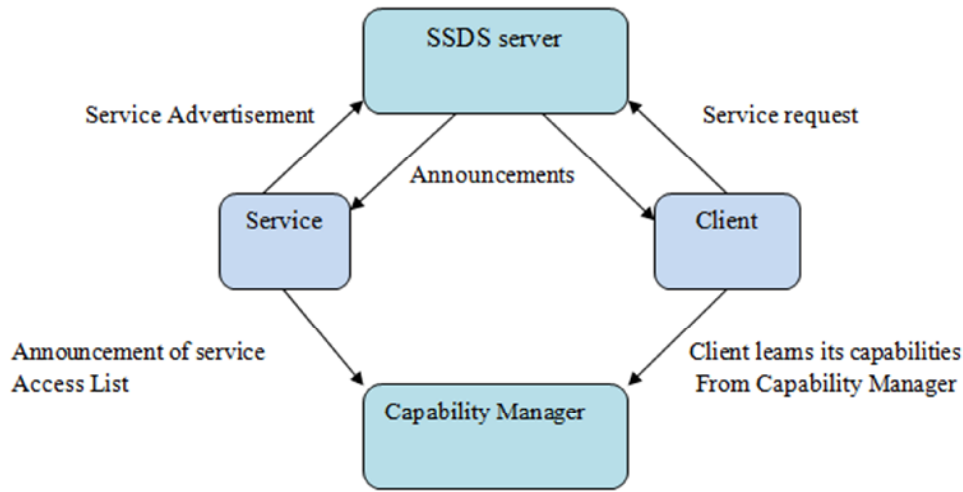

Figure-7- SSDS Architecture

In SSDS, the service discovery is handled as follows:

SSDS servers periodically announce their existence on a globally known multicast channel. These announcements include the necessary information to contact SDS server, CA, and CM. Services and clients listening for can join the system. A service can advertise itself by registering its service description with the SSDS server. A typical service description is expressed in XML and includes service location, service lifetime, and other descriptive attributes. Queries discovery are also expressed using XML and assessment is achieved by the SSDS server by comparing XML tags of queries with those included in service description. SSDS uses Authenticated RMI (Remote Invocation Method) for all communications.

Regarding Configuration Update, SSDS does not implement the mechanism of leasing or notification to control the availability of a service. During the announcement process, the service indicates to SSDS server period of time for which the service is still valid. After the expiry of this period, the service is considered unavailable by the server SSDS.

## 3. Comparison:

Service Discovery Protocols discussed above in this paper are compared taking various criteria as service description, architecture, search filter, service usage, operation without directory, configuration purge, programming language, and interoperability between protocols. Note that Bluetooth SDP may be omitted in some parts of this comparison. This is mainly due to its limited functionality (addresses only searching or browsing for service). A summary of the comparisons is presented in **Appendix A**.

### 3.1. Service Description:

Services first of all, need to be described in term of their main functions and capabilities before getting involved into service discovery procedure. All the aforementioned protocols adopt the Key-Value approach for service description using various formats (java object and assistant attribute in Jini, XML in UPnP...). The key-value approach is much simple technique which helps users to make more informed decisions upon selecting among similar services. On the other side, it implies that service providers and users have to agree on the exact keywords before the service can be recognized while different key- words can offer the same functionality. This conflict can be solved if an ontology has been adopted, so that semantic matching is possible and keyword similarity can be taken into account when searching for services [13].

In future, heterogeneity will occur in various aspects (hardware, software platforms, network protocols….) so using a common format for service description will be essential to preserve interoperability between service discovery protocols and heterogeneous nodes involved in the discovery process.

## 3.2. Service Discovery Architecture:

Depending on how service descriptions are stored, architecture adopted by service discovery protocols can be classified into two models:

***Directory based architecture***: adopted by SLP, Jini, Salutation, and SSDS. This architecture is more suited for large network since a directory increases performance and facilitate discovery. However, using central directory can make the system susceptible to a single point of failure [6]. One partial solution, is to deploy a distributed directory as Jini does (the information is portioned and stored at different LSs) therefore the failure of one LS leads to unavailability to a part of the directory.

***Non-directory based model***: adopted by UPnP and SDP. This architecture does not require any directory; the information is always up-to-date since services respond directly to queries. However, this architecture does not scale and becomes less attractive when number of services growth in the network [4] due to the extensive use of multicast and broadcast for communications.

## 3.3. Operation without central directory:

Except Jini which needs its lookup Service in order to operate and SSDS which requires the presence of the SSDS server, the other architectures can manage without directory. SLP and Salutation provide more flexible solution, since they support both architectures. This feature allows SLP and Salutation to operate in different environments and taking the advantages of both architectures.

## 3.4 Service Matching (Discovery Filter)

Service matching is also an important component in service discovery, as it determines the relevance of the service on the user's request and then concludes whether it should be returned as a result. The effectiveness of service matching depends on the expressiveness of the information needs (query). Regarding the surveyed protocols, when a query language is adopted, the query is more expressive and attributes can be compared using operators and comparators (that is SLP and Salutation). Otherwise, evaluation of requests is based on exact correspondence between request parameters and service type and attributes (Jini, SSDS, UPnP, and SDP). Note that UPnP supports only service type matching [16].

## 3.5. Service Invocation:

After the discovery stage, the user invokes selected service. Service invocation involves the service network address and an underlying communication mechanism [4] between the client and invoked service. Aforementioned service discovery protocols provide two different support levels:

At the first level, protocols provide only service location (that is SLP). Applications are responsible for defining the communication mechanism. At the second level, in addition to service location, a protocol defines the underlying communication mechanisms. For example: in Salutation communications can be done through RPC. In Jini and SSDS, clients use java Remote Method Invocation, and UPnP uses SOAP protocol based on XML and HTTP to control services.

### 3.6. Configuration Update (management dynamicity)

All presented service discovery protocols allow automatic detection of disconnected services using leasing or TTL except Salutation, where the user has explicitly to ask its local SLM to check availability of the desired service.
Jini and UPnP implement also an event notification system to inform users about service state changes. In this case, users have to register their interest in receiving service state updates.

### 3.7. Programming Language and Network independence:

All of the examined approaches are designed for IP networks and do not depend on any programming language, except Jini and SSDS which are implemented with java. Note that Salutation is transport independent and can operate over different network not only in IP network. This transport independence is the strongest feature of Salutation.

### 3.8. Interoperability between service discovery protocols

All the examined protocols do not interoperate with each other due to the variety of service description formats and communication protocols. UPnP on his side uses standard technologies such as: XML, HTTP, and SOAP which make easy for other discovery mechanisms to interwork with it.

There is a number of works bridging Jini-UPnP [19], SDP-UPnP [20], Jini-SLP [21], and Salutation-SDP [22]. It is beyond the scope of this article to analyze approaches proposed for service discovery protocol interoperability. However, one can judge that these works still remain partial solutions and try to enhance interworking between protocols by only implementing of certain low-level function of service discovery [3]. This is mainly due to some mappings simply are not achievable (e.g. UPnP eventing cannot be mapped to any SLP function or SDP function).

## 4. Conclusion

This paper surveys five of the leading technologies in service discovery. In particular we clarify service discovery features and techniques, so that the reader can have an overall picture of the service discovery system. We discussed the fundamental architectures for service discovery, explaining the basic ideas for each architecture, and describing how each one ensure the functionalities of an autonomic service discovery protocol. Then, we make a comparison between them and commenting the advantages and disadvantages for each one. This comparison is based on eight prime criteria for autonomic service discovery, which we have defined before.

As a conclusion, all the surveyed protocols constitute a good solution for service discovery. There is no clear dominance of one solution over the others, since each one has its advantages and drawbacks. By analyzing the characteristics of examined approaches, a future service discovery protocols should have the following characteristics:

- **Ontology for service description:** Service description should define the functionality and intention of a service in unambiguous way. This can be accomplished only if an ontology has been adopted, so that semantic matching is possible and keyword similarity can be taken into account when searching for services.
- **Context awareness**: Another important consideration for service discovery is context awareness. It plays key feature for providing more suitable services to clients by taking into account different information in the discovery stage (such as: user preference, terminal capabilities, QoS…).
- **Service selection:**  after submitting a query for certain service, there may be several services meet the criteria. Service discovery protocol should incorporate an automatic service selection mechanism based on a set of metrics to define the best service offer.
- **Robustness:** A discovery mechanism should be capable to cope with faults and network changes without considerable performance or function losses. An example is recovery from errors or the structural re-stabilization of multiple sites.
- **Interoperability:** Considering the multitude of service discovery standards, architectures and protocols and taking also into account the heterogeneity nature of future environments, interoperability in service discovery will be a major issue requiring attention.

**Authors**

**Bendaoud Karim Talal** received his Master degree in Telecommunication Engineering from the University of Tlemcen (Algeria) in 2011. Member of STIC laboratory in the same university, his research interests include smart spaces

**Merzougui Rachid** received his PhD in Telecommunication Engineering from the University of Tlemcen (Algeria) in 2012. His research interests cover telecommunication systems and mobile networks.

**Appendix A -** Comparative Table on Service Discovery Protocols

| Criteria | SLP | JINI | Bluetooth SDP |
|---|---|---|---|
| Mains entities | User Agent Service Agent Directory Agent | Lookup Service Service ,Client | Service Client |
| Architecture | both architectures (can be Directory or Non- directory) | Directory based | Non-directory |
| Service Description | Service URL + Service Template | Java Object (proxy) + Attributes | Service Record |
| Storage for Service Description | On Directory Agent / On every SA | On Lookup Service | on every SDP server |
| Service Announcement | Registration with DA / Multicast Announcement | Registration with Lookup service | -- |
| Service Discovery | Query the DA/ Multicast to SAs | Query the Lookup Service | Query the SDP server |
| Service Matching | Powerful | Exact correspondence with service attributes | Exact correspondence with service attributes |
| Operation without Directory | Yes | Lookup Service required | Yes |
| Service Usage | Provide only service location | Java RMI via proxy Object | -- |
| Configuration Update | Service registration life time | Leasing + remote events | -- |
| Programming Language &Network independence | Independent | Java | independent |
| | IP | IP | independent |
| Interoperability | Possible, e.g. Jini, | Possible, e.g., UPnP. | Possible, e.g., Salutation |

| Criteria | Salutation | UPnP | SSDS |
|---|---|---|---|
| Mains entities | Salutation Manager Transport Manager Client and Service | Devices Control points | SDS server Certificate Authority Capability Manager |
| Architecture | both architectures (can be Directory or Non- directory) | Non-directory | Directory based |
| Service Description | Functional Unit Service Record | XML for description | XML for description |
| Storage for Service Description | Service Registry on every SLM | On every UPnP device | SSDS required |
| Service Announcement | Registration with local SLM | Multicast advertisement (ssdp:alive) | On every SSDS server |
| Service Discovery | Query the local SLM | Listen to advertisement / Multicast the devices | Query the SSDS server |
| Service Matching | Powerful | Limited to service type or ID | Exact correspondence with service attributes |
| Operation without Directory | Yes | Yes | SSDS server required |
| Service Usage | Usage through RPC | SOAP Protocol | Usage through RMI |
| Configuration Update | Check availability with local SLM | Advertisement life time + event notification | Service registration life time |
| Programming Language and Network independence | Independent | Independent | Java |
| | Independent | IP | IP |
| Interoperability | Possible, e.g., SDP | Possible, e.g., Jini and SDP | -- |